\documentclass[aps,prc,showpacs,showkeys,twoside,twocolumn,byrevtex,floatfix]{revtex4-2}

\usepackage[dvips]{graphicx} % PRC
\usepackage{color}           % PRC % for color  usage: \textcolor{red}{text}
\usepackage{amsmath,amssymb,amsfonts}
\usepackage{bm}
\usepackage{mathtools}
\newcommand{\nc}{\newcommand}           % new command
\nc{\vc}[1]     {\mbox{\boldmath $#1$}} % boldmath(vector)
\nc{\mapleft}[1]{                       % something under arrow
 \smash{\mathop{                      %
  \hbox to 0.90cm{\rightarrowfill} }\limits_{#1}}}

\nc{\beq}     {\begin{eqnarray}}
\nc{\eeq}    {\end{eqnarray}}
\nc{\bra}       {\langle}               % bra
\nc{\ket}       {\rangle}               % ket
\nc{\bras}[1]   {\langle#1|}            % <#1|
\nc{\kets}[1]   {|#1\rangle}            % |#1>
\nc{\del}       {\partial}              % bra state
\nc{\wtil}      {\widetilde}            % wide-tilde
\nc{\hO}        {\hat{O}}           % operator
\nc{\EV}[1]     {\langle #1 \rangle} % expectation value

\newcommand{\lw}[1]{\smash{\lower1.75ex\hbox{#1}}}

\nc{\mydraft}	{\setlength{\topmargin}{-1.5cm}}
\mydraft
\begin{document}

\title{Generalized coherent states satisfying the Pauli principle in a nuclear cluster model}

\author{Takayuki Myo}
\email{takayuki.myo@oit.ac.jp}
\affiliation{General Education, Faculty of Engineering, Osaka Institute of Technology, Osaka, Osaka 535-8585, Japan}
\affiliation{Research Center for Nuclear Physics (RCNP), Osaka University, Ibaraki, Osaka 567-0047, Japan}

\author{Kiyoshi Kat\=o}
\email{kato@nucl.sci.hokudai.ac.jp}
\affiliation{Nuclear Reaction Data Centre, Faculty of Science, Hokkaido University, Sapporo 060-0810, Japan}

\date{\today}

\begin{abstract}%
  We propose a new basis state, which satisfies the Pauli principle in the nuclear cluster model.
  The basis state is defined as the generalized coherent state of the harmonic oscillator wave function
  using a pair of the creation operators and is orthogonal to the Pauli-forbidden states having smaller quanta.
  In the coherent basis state, the range parameter is changeable and controls the radial dilation.
  This property is utilized for the precise description of the relative motion between nuclear clusters.
  We show the reliability of this framework for the $2\alpha$ system of $^8$Be in the semi-microscopic orthogonality condition model.
  We obtain the resonances and non-resonant continuum states of $2\alpha$ with complex scaling. 
  The resonance solutions and the phase shifts of the $\alpha$-$\alpha$ scattering agree
  with those using the conventional projection operator method to remove the Pauli-forbidden states.
  We further discuss the extension of the present framework to the multi-$\alpha$ cluster systems using the SU(3) wave functions.
\end{abstract}

\pacs{
21.60.Gx, % Cluster Models
27.20.+n~ % 6(less-than-or-equal-to)A(less-than-or-equal-to)19
}
\maketitle

%%%%%%%%%%%%%%%%%%%%%%%%%%%%%
\section{Introduction}\label{sec:intro}
Nuclear clustering is a fundamental aspect of nuclei \cite{ikeda68,horiuchi12,freer18}, such as the spatial formation of $\alpha$ clusters in nuclei.
The $^8$Be nucleus is a typical cluster system decaying into two $\alpha$ particles.
In $^{12}$C, the $0^+_2$ state is known as the Hoyle state having a three $\alpha$-structure
located near the three $\alpha$ threshold energy.

In nuclear cluster models, the resonating group method (RGM) \cite{wheeler37,horiuchi70} is a microscopic approach
starting from the degrees of freedom of nucleons and used to solve the relative motions between clusters in nuclei.
The orthogonality condition model (OCM)  \cite{saito68} is a semi-microscopic approach,
in which the local potential is often used as the intercluster potential to fit the experimental data of the cluster systems.
This is the advantage of OCM to reproduce the threshold energies of every cluster emission in nuclei.

The Pauli principle is an essential statistics of nuclei and this property is fully treated in RGM.
The Pauli-forbidden states are defined as the zero-eigenvalue states of the RGM norm kernel.
In OCM, the Pauli-forbidden states are removed from the space of relative motion between clusters,
and only the Pauli-allowed states are treated and obtained dynamically. 

When the nuclear clusters are described with the harmonic oscillator (HO) shell model wave functions,
the Pauli-forbidden states are also expressed by using the HO states for the relative wave function between clusters.
Technically there are several methods to remove the Pauli-forbidden states in relative motion in OCM.
One is the Gram-Schmidt orthonormalization method \cite{kruppa90}.
When the relative motion is precisely solved by using the linear combination of the appropriate basis functions,
Kukulin's projection operator method works to push the Pauli-forbidden states in every relative motion to the irrelevant energy region \cite{kukulin78}.
In this method, the pseudo potential with the projection operator form to the Pauli-forbidden states is added to
the Hamiltonian and the orthogonal solutions can be obtained as physical states.
This method sometimes makes difficulty increasing the number of clusters in multicluster systems such as $4\alpha$ and $5\alpha$,
because precise projections are necessary for every cluster-pair to eliminate the Pauli-forbidden states in all the relative motions,
which causes the numerical efforts with many basis states.
In this situation, one needs an efficient method to treat the Pauli-allowed states in the description of multicluster systems based on OCM.

In this paper, we propose a new scheme to treat the Pauli-allowed states in OCM;
all the basis states in relative motion are automatically orthogonal to the Pauli-forbidden states
and it is not necessary to use the projection operator in the Hamiltonian and the wave function.
We formulate this method in the generalized coherent states \cite{perelomov86} of the HO basis states for relative motion between clusters
using the raising operator $\hat{\vc{a}}^\dagger \cdot \hat{\vc{a}}^\dagger$ \cite{rowe85,yoshida11}.
This operator increases the quanta of every HO state and can be utilized to define the Pauli-allowed states above the Pauli-forbidden states. 
In this method, we can describe the resonances in cluster-cluster scattering using the complex scaling \cite{myo14}.

In this paper, we formulate the new method and confirm its reliability by calculating the $2\alpha$ system of $^8$Be,
in which we use the complex-scaled solutions of the resonant and nonresonant continuum states.
The present work becomes the foundation to investigate the multicluster systems in the OCM approach.

In Sec.~\ref{sec:method}, we provide the formulation of the generalized coherent state with the HO basis states
and its application to the nuclear cluster systems.
In Sec.~\ref{sec:result}, we discuss the resonances and scattering of the 2$\alpha$ system of $^8$Be.
In Sec.~\ref{sec:discuss}, we discuss the extension to the multi-$\alpha$ cluster systems using the SU(3) wave functions.
In Sec.~\ref{sec:summary}, we summarize this work.
In the Appendix, we give the mathematical derivation of the generalized coherent states with the HO basis states.

%%%%%%%%%%%%%%%%%%%%%%%%%%%%%
\section{Theoretical methods}\label{sec:method}

%%%%%%%%%%%%%%%%%%%%%%%%%%%%%%%%%%%%%%%%%%%%%%%%%%%%%%%%%%%%%%%%%%%%%%%%%%%%%%%%%%%%%%%%
\subsection{Generalized coherent states}\label{sec:PAS}
We begin with the harmonic oscillator (HO) basis state $\phi_{n\ell m}(\vc{r},\nu)$ with a range  $\nu=1/b^2$ and a principal quantum number $N=2n+\ell$,
where $n$ represents the number of nodes in the radial wave function and $\ell$ is an orbital angular momentum.
Using the operators of the creation and annihilation of a quanta $N$, the HO basis state can be written as \cite{bergmann60,moshinsky96} 
\begin{equation}
  \begin{split}
    \phi_{n\ell m}(\vc{r},\nu) &= A_{n\ell}\, \left( \hat{\vc{a}}^\dagger \cdot \hat{\vc{a}}^\dagger \right)^n
        {\cal Y}_{\ell m} (\hat{\vc{a}}^\dagger)\, \phi_{0}(\vc{r},\nu),
  \\
  A_{n\ell}&= (-1)^n \sqrt{\frac{4\pi}{(2n+2\ell+1)!!\, (2n)!!}},
  \\
  \phi_{0}(\vc{r},\nu) &=\left(\frac{\nu}{\pi}\right)^{3/4} e^{-\frac12 \nu r^2},
  \label{eq:HO_aa}
  \end{split}
\end{equation}
where ${\cal Y}_{\ell m} (\vc{r})=r^\ell Y_{\ell m} (\hat{\vc{r}})$ is a solid spherical harmonics and $\phi_{0}$ is a vacuum with $N=0$. 
This HO basis state can be used to represent the single-nucleon wave function in nuclei and also the relative wave function between nuclear clusters.

In this study, we introduce the following scalar operators $\hat{D}^\dagger$ (raising) and $\hat{D}$ (lowering) following Ref. \cite{rowe85} as 
\begin{equation}
  \begin{split}
    \hat{D}^\dagger&= \hat{\vc{a}}^\dagger \cdot \hat{\vc{a}}^\dagger,\qquad
    \hat{D} = \hat{\vc{a}} \cdot \hat{\vc{a}}.
  \end{split}
\end{equation}
These operators belong to the symplectic Sp(3,$\vc{R}$) Lie algebra of the coherent state of the collective motion
and change the quanta of the wave function by two for the radial part without changing the angular momentum.
Using the raising operator $\hat{D}^\dagger$, we introduce the following new basis state $\phi_{n \ell m}^\beta(\vc{r},\nu)$,
in which $\hat{D}^\dagger$ is coherently multiplied by the HO basis state with the weight of the real parameter $\beta$ in the exponential form as
\begin{equation}
  \begin{split}
   \phi_{n \ell m}^\beta(\vc{r},\nu)
   &=\exp\left(\frac12 \beta \hat{D}^\dagger \right)\, \phi_{n \ell m}(\vc{r},\nu)
   \\
   &=\frac{1}{\sqrt{(1+\beta)^{N+3/2}}}\, \exp\left( \frac{\beta}{2(1+\beta)} \nu r^2 \right)\,
   \\
   &\times \phi_{n \ell m}(\vc{r},\frac{\nu}{1+\beta}).
   \label{eq:coherent}
  \end{split}
\end{equation}
The derivation of this equation is given in Appendix \ref{sec:appendix_A}.
This new basis state is a kind of generalized coherent state  \cite{perelomov86} in terms of $\hat{D}^\dagger$
and can be represented by the HO basis state with the same quanta $N$ and the different range parameter $\nu/(1+\beta)$
and multiplying the Gaussian function with the coordinate $r$. This equation plays an essential role in this study.
It is noted that the basis state has the following exponential dependence;
\begin{equation}
  \begin{split}
   \phi_{n \ell m}^\beta(\vc{r},\nu)&\propto   \exp \left(- \frac{1-\beta}{2(1+\beta)}\,  \nu r^2 \right).
   \label{eq:HO_exp}
  \end{split}
\end{equation}
This form gives a condition of $|\beta|<1$ to satisfy the asymptotically damping behavior, which is imposed throughout this study.
When $n=0$, the basis state $\phi_{0 \ell m}^\beta(\vc{r},\nu)$ becomes the nodeless Gaussian function multiplying $r^\ell$,
which is often used in the Gaussian expansion technique \cite{kamimura88,kameyama89,aoyama06}.

From the property of the raising operator $\hat D^\dagger$,
the function $\phi_{n \ell m}^\beta(\vc{r},\nu)$ includes only the quanta larger than or equal to $N$ of the HO basis states with the range $\nu$.
Hence the following orthogonal condition is satisfied;
\begin{equation}
  \langle \phi_{n' \ell m}(\nu) | \phi_{n \ell m}^\beta(\nu) \rangle = 0 \qquad\mbox{for}~n'<n~(N'<N).
\end{equation}
This property is useful to construct the HO basis states with a quanta $N$ and any values of $\beta$, which are orthogonal to the HO states with a lower quanta $N'$.
If one regards the HO basis states with the lower quanta $N'$ as the occupied states in the nucleus, namely the Pauli-forbidden states,
the generalized coherent basis states $\phi_{n \ell m}^\beta(\vc{r},\nu)$ can be the unoccupied states in the nucleus automatically,
and represents the Pauli-allowed states.
A specific case of this formulation is introduced in the shell model,
in which the HO particle state with a free range parameter is taken to be orthogonal to the HO hole states
by adjusting the polynomial in the HO particle state \cite{myo07,myo11}.

The parameter $\beta$ controls the spatial range of the generalized coherent basis states.
When $\beta$ is close to unity, the basis state has a long tail and is suitable to describe a weak-binding state of nuclei such as a halo structure and the low-energy scattering solution in the nuclear reaction. 
When $\beta$ is close to $-1$, the basis state becomes a short-range and is suitable to describe the short-range and tensor correlations of nucleons with high momenta in nuclei \cite{myo07}.
From these properties, the parameter $\beta$ plays a role on the radial dilation of the coherent basis state,
and then we call $\beta$ ``dilation parameter'' hereafter.

In the cluster model, the present coherent basis state is useful to describe
relative motion between clusters with the orthogonality condition from the Pauli principle for the following two reasons:
\begin{enumerate}
\item[(i)]
  When the cluster wave functions are the HO shell-model ones, the Pauli-forbidden states in relative motion become the HO states with a specific quanta $N_{\rm PF}$.
  Hence, the coherent basis states with a relative oscillator quanta $N$ and $\beta$ become the Pauli-allowed states that are orthogonal to the Pauli-forbidden states
  with the condition of $N_{\rm PF}<N$ \cite{kato19}.
\item[(ii)]
  The relative motion between clusters is solved precisely
  and the relative wave function is optimized by superposing the coherent basis states $\phi_{n \ell m}^\beta(\vc{r},\nu)$
  with various dilation parameters $\beta$,
  each of which shows a different spatial distribution.
\end{enumerate}

In the multicluster system, we can prepare the cluster wave function using the coherent basis states in relative motion between every cluster-pair.
In this paper, we consider the two-cluster case with clusters C$_1$ and C$_2$
and one intercluster motion with the coordinate $\vc{r}$ in the single channel.
We express the total nuclear wave function $\Psi$, in which the relative wave function $\Phi_{\rm rel}(\vc{r})$ is in the linear combination form
of the coherent basis states $\{\phi^{\beta_i}_{n \ell m}(\vc{r},\nu_{\rm rel})\}$ with the range parameter $\nu_{\rm rel}$,
the set of $\{\beta_i\}$ with $i=1,\cdots,N_{\rm base}$, and 
the condition of $N=2n+\ell$ for Pauli-allowed states, $N > N_{\rm PF}$;
\begin{equation}
\begin{split}
  \Psi &= {\cal A}\{\phi_{{\rm C}_1}\phi_{{\rm C}_2}\Phi_{\rm rel}(\vc{r})\},
  \\
  \Phi_{\rm rel}(\vc{r}) &=   \sum_{i=1}^{N_{\rm base}} C_i\, \phi^{\beta_i}_{n \ell m} (\vc{r},\nu_{\rm rel}),
  \label{eq:WF_2alpha}
\end{split}
\end{equation}
where ${\cal A}$ is the antisymmetrizer of nucleons between different clusters and $\phi_{{\rm C}}$ is the internal wave function of the cluster $\rm C$.
Hereafter we omit the notation of the quantum numbers $n$,$\ell$, and $m$ in the basis states for simplicity.
It is possible to add the basis states with different $n$ to $\Phi_{\rm rel}(\vc{r})$ as well as $\beta_i$.

In the present study, we adopt the orthogonality condition model (OCM).
The eigenvalue problem of the Hamiltonian $H$ for relative motion is given to obtain the relative energy $E$ between clusters:
\begin{equation}
\begin{split}
H&=T_{\rm rel} + V_{{\rm C}_1{\rm C}_2},
\\
H\Phi_{\rm rel}(\vc{r}) &=E \Phi_{\rm rel}(\vc{r}) ,
\\
\sum_{j=1}^{N_{\rm base}} (H_{ij}& -E N_{ij})\, C_j=0,
\\
H_{ij}&=\langle \phi^{\beta_i}(\nu_{\rm rel})|H|\phi^{\beta_j}(\nu_{\rm rel}) \rangle,
\\
N_{ij}&=\langle \phi^{\beta_i}(\nu_{\rm rel}) |\phi^{\beta_j}(\nu_{\rm rel}) \rangle,
\label{eq:coherent_eigen}
\end{split}
\end{equation}
where $T_{\rm rel}$ and $V_{{\rm C}_1{\rm C}_2}$ are the kinetic energy and the potential of relative motion between clusters, respectively.
The matrix elements of $H_{ij}$ and $N_{ij}$ are those of the Hamiltonian and norm with the individual $\beta$-values, respectively.
In this paper, we call the present framework ``coherent basis method''.

In the coherent basis method, the matrix elements can be calculated analytically,
and we use the formulas using the HO basis states with the independent range parameters
in the bra and ket states \cite{kruppa90}, characterized by $\beta_i$ and $\beta_j$ in Eq. (\ref{eq:coherent_eigen}).
For kinetic energy, we give the formula in Appendix \ref{sec:appendix_B}. 
\subsection{$\alpha$-$\alpha$ system}

We demonstrate the present new scheme in the $\alpha$-$\alpha$ cluster system of $^8$Be.
The $\alpha$ cluster is represented by the $(0s)^4$ configuration of the HO basis state
where the range parameter $\nu$ of the single-nucleon state is taken as 0.535 fm$^{-2}$, which corresponds to the length of $b=1.3672$ fm, to reproduce the charge radius of the $\alpha$ particle.
We prepare the coherent basis states for the relative wave function of $2\alpha$,
the range parameter of which is $\nu_{\rm rel}=2\nu=$ 1.070 fm$^{-2}$ corresponding to the length of $b_{\rm rel}=0.9667$ fm.
We employ the folding potential between $\alpha$-$\alpha$ with the nucleon-nucleon interaction and the Coulomb interaction using the $\alpha$ cluster wave function.
We adopt the Schmid-Wildermuth effective nucleon-nucleon interaction \cite{schmid61}, which is often used in the previous studies of the multi-$\alpha$ cluster systems \cite{fukatsu92,kurokawa05,kurokawa07,funaki08}.
The form of the $\alpha$-$\alpha$ folding potential $V_{\alpha\alpha}(r)$ is given with nuclear (N) and Coulomb (C) parts as
\begin{equation}
\begin{split}
  V_{\alpha\alpha}(r)&=V^{\rm N}_{\alpha\alpha}(r)+V^{\rm C}_{\alpha\alpha}(r),
  \\
  V^{\rm N}_{\alpha\alpha}(r)
  &= 2\, X_D\, V_0 \, a^{3/2} e^{-a \mu r^2},
  \\
  X_D &= 8W+4B-4H-2M,\quad
  a=\frac{2\nu}{2\nu+3\mu}
  \\
  V^{\rm C}_{\alpha\alpha}(r)&= 4\, e^2\, \frac{{\rm erf}(cr)}{r},\qquad
  c=\sqrt{\frac{2\nu}{3}},
\end{split}
\end{equation}
where $r=|\vc{r}|$, $V_0=-72.98$ MeV, $\mu= 0.46$ fm$^{-2}$, $W=M=0.4075$, and $B=H=0.0925$.

The lowest shell-model configuration of the $2\alpha$ system is $(0s)^4(0p)^4$ in the HO basis state with a total quanta of 4.
Hence the Pauli-forbidden HO states, $\omega_{\rm F}(\vc{r},\nu_{\rm rel})$, are defined
by the condition of quanta $N_{\rm PF} < 4$ in the relative motion with the range $\nu_{\rm rel}$.
In the coherent basis state, we impose this condition of the Pauli-allowed states and
set $N=4$ for the $0^+$, $2^+$, and $4^+$ states and $N=6$ for the $6^+$ state in the present study.
For the $4^+$ and $6^+$ states, there is no Pauli-forbidden state.

We take various dilation parameters $\beta_i$ in Eq.~(\ref{eq:WF_2alpha}) to optimize the radial wave function.
In the present study, we choose the set of $\beta_i$ in the form of the geometric progression
of the length parameters $b_i$ of the HO basis state \cite{kamimura88,kameyama89} according to Eq. (\ref{eq:HO_exp}) as
\begin{equation}
\begin{split}
  \frac{1-\beta_i}{1+\beta_i}\, \nu_{\rm rel} & =\frac{1}{b_i^2} =\frac{1}{(b_0\gamma^{i-1})^2}.
  \label{eq:beta-b}
\end{split}
\end{equation}
We set $b_0 =0.2$ fm, $\gamma=1.2$, and $N_{\rm base}=30$ in the present calculation, which are transformed to $\beta_i$ in the coherent basis states.

To show the reliability of the coherent basis method,
we compare the obtained results with those of the conventional projection operator method (PO) \cite{kukulin78}.
In PO, one usually adds the pseudo potential of the projection operators with a positive prefactor $\lambda$ to the original Hamiltonian given as;
\begin{equation}
\begin{split}
  H_\lambda &= H + \lambda \sum_{f} |\omega_{{\rm F},f}\rangle \langle \omega_{{\rm F},f}|.
  \label{eq:PSE}
\end{split}
\end{equation}
One uses a large value of $\lambda$ to make the solutions orthogonal to the Pauli-forbidden states $\{\omega_{{\rm F},f}\}$ with the index $f$
and we take $\lambda=10^6$ MeV in this study \cite{aoyama06}.
The number of the Pauli-forbidden states is determined from the condition of of quanta as $N_{\rm PF}=2n_f+\ell_f<4$.
For the basis states of the relative motion in PO, we adopt the nodeless HO basis functions with $n=0$, which are often used in the OCM calculation, as
\begin{equation}
\begin{split}
  \Phi_{\rm rel}(\vc{r}) &=   \sum_{i=1}^{N_{\rm base}} \bar C_i\, \phi_{0 \ell m }(\vc{r},b_i),
\\
  \phi_{0 \ell m }(\vc{r},b_i)&= N_\ell(b_i)\, e^{-1/2(r/b_i)^2}\, {\cal Y}_{\ell m}(\vc{r}),
  \label{eq:WF_PO}
\end{split}
\end{equation}
where $N_\ell(b)$ is a normalization factor of the basis state.
The choice of the length parameters $\{b_i\}$ is the same as those of the coherent basis states
in Eq.~(\ref{eq:beta-b}), which is suitable for  comparing the solutions.

%
%%%%%%%%%%%%%%%%%%%%%%%%%%%%%%%%%%%%%%%%%%%%%%%%%%%%%%%%%%%%%%%%%%%%%%%%%%%%%%%%%%%%%%%%
\subsection{Complex scaling}\label{sec:csm}

We describe the resonances and the scattering of the $\alpha$-$\alpha$ system using the complex scaling \cite{myo14,aoyama06,ho83,moiseyev98,myo20}
in both the coherent basis method and the projection operator method.
In the complex scaling, the relative coordinate $\vc{r}$, the relative momentum $\vc{p}$ in the Hamiltonian $H$,
and the relative wave function $\Phi_{\rm rel}(\vc{r})$ are transformed using a scaling angle $\theta$ with an operator $U(\theta)$ as
\begin{equation}
U(\theta)~:~\vc{r} \to \vc{r}\, e^{ i\theta},\qquad
\vc{p} \to \vc{p}\, e^{-i\theta} ,
\label{eq:CSM}
\end{equation}
where $\theta$ is a real positive number. 
The complex-scaled Hamiltonian $H^\theta$, the complex-scaled relative wave function $\Phi_{\rm rel}^\theta$,
and the corresponding energy $E^\theta$ are given as
\begin{equation}
\begin{split}
  H^\theta&=  U(\theta)H U^{-1}(\theta) ,
  \\
  \Phi_{\rm rel}^\theta (\vc{r})
  &= U(\theta)\Phi_{\rm rel}(\vc{r})
  \\
  H^\theta \Phi_{\rm rel}^\theta (\vc{r}) &= E^\theta \Phi_{\rm rel}^\theta(\vc{r}).
  \label{eq:WF}
\end{split}
\end{equation}

After solving the last equation, $E^\theta$ are obtained for bound, resonant, and continuum states on the complex energy plane according to the ABC theorem \cite{ABC}.
The energies of the continuum states start from the $\alpha$+$\alpha$ threshold energy and are obtained along the line rotated down by $2\theta$ from the real energy axis.
The energies of the bound and resonant states are independent of $\theta$.
The resonance has a complex energy $E_{\rm R}=E_r-i\Gamma/2$ with a resonance energy $E_r$ and a decay width $\Gamma$.
The asymptotic behavior of the resonance wave function becomes a damping form if $2\theta > |\arg(E_{\rm R})|$ \cite{ABC}.
In calculations with a finite number of the basis states,
the resonances are identified from the stationary property of $E_{\rm R}$ with respect to $\theta$ on the complex energy plane \cite{aoyama06,ho83,moiseyev98},
and the continuum states are discretized with the complex energies.
The wave function $\tilde \Phi_{\rm rel}^\theta (\vc{r}) $ is the biorthogonal state of $\Phi_{\rm rel}^\theta (\vc{r}) $ \cite{berggren68},
and used for the bra state in the complex-scaled matrix elements.
One does not take the complex conjugate of the radial part of the bra state in the matrix elements \cite{ho83,moiseyev98}.

The Pauli forbidden state $\omega_{\rm F}(\vc{r},\nu_{\rm rel})$ is also transformed in the complex scaling as
\begin{equation}
\begin{split}
  U(\theta)\omega_{\rm F}(\vc{r},\nu_{\rm rel})
  &= \omega_{\rm F}^\theta(\vc{r},\nu_{\rm rel}) = e^{3i\theta/2}\omega_{\rm F}(\vc{r} e^{i\theta},\nu_{\rm rel})
  \\
  &= \omega_{\rm F}(\vc{r},\nu_{\rm rel} e^{2i\theta}).
\end{split}
\end{equation}
In the last equation, we use the explicit form of the HO basis state, and the range parameter $\nu_{\rm rel}$ is transformed
instead of $\vc{r}$.

In the projection operator method, the Hamiltonian $H_\lambda$ in Eq.~(\ref{eq:PSE}) is transformed as
$H_\lambda^\theta =  U(\theta)H_\lambda U^{-1}(\theta)$, in which the Pauli-forbidden states are transformed in the pseudo potential \cite{aoyama06}.
In analogy with Eq.~(\ref{eq:WF_PO}), the complex-scaled wave function is given as 
\begin{equation}
\begin{split}
  \Phi^\theta_{\rm rel}(\vc{r}) &= \sum_{i=1}^{N_{\rm base}} \bar C^\theta_i\, \phi_{0 \ell m }(\vc{r},b_i),
\end{split}
\end{equation}
where the $\theta$ dependence is included in $\{\bar C^\theta_i\}$.
This expansion is often used in the conventional OCM calculation with complex scaling \cite{aoyama06}.

In the coherent basis method, the coherent basis state with a dilation parameter $\beta$ in Eq.~(\ref{eq:WF_2alpha})
is transformed because the basis state should be orthogonal to the complex-scaled Pauli-forbidden states as
\begin{equation}
  \langle \tilde \omega_{\rm F}^\theta(\nu_{\rm rel}) | \phi^{\beta,\theta}(\nu_{\rm rel}) \rangle =
  \langle \tilde \omega_{\rm F}(\nu_{\rm rel} e^{2i\theta}) | \phi^{\beta}(\nu_{\rm rel} e^{2i\theta}) \rangle = 0.
\end{equation}
Hence the relative wave function $\Phi_{\rm rel}^\theta (\vc{r})$ is expanded in terms of
the complex-scaled coherent basis states $\{ \phi^{\beta_i,\theta}(\vc{r},\nu_{\rm rel}) \}$ with the index $i$ for $\beta_i$ as
\begin{equation}
\begin{split}
  \Phi_{\rm rel}^\theta (\vc{r})
  &= \sum_{i=1}^{N_{\rm base}} C_i^\theta \phi^{\beta_i,\theta}(\vc{r},\nu_{\rm rel}) ,
\end{split}
\end{equation}

One solves the following eigenvalue problem of the complex-scaled Hamiltonian matrix and obtains $E^\theta$ and $\{C_i^\theta\}$ for each eigenstate; 
\begin{equation}
   \sum_{i=1}^{N_{\rm base}} \left( H_{ij}^\theta - E^\theta N_{ij}^\theta \right) C_j^\theta = 0.
   \label{eq:eigen}
\end{equation}

Technically, the matrix elements with the complex-scaled coherent basis states are calculated in the following procedure;
\begin{equation}
\begin{split}
H_{ij}^\theta &=\langle \tilde \phi^{\beta_i,\theta} (\nu_{\rm rel}) |H^\theta|\phi^{\beta_j,\theta} (\nu_{\rm rel}) \rangle
\\
&=\sum_{p,q}^{N_p} \langle \tilde \phi^{\beta_i,\theta}(\nu_{\rm rel})| \phi_p \rangle \, \langle  \tilde  \phi_p | H^\theta| \phi_q \rangle \, \langle  \tilde  \phi_q |\phi^{\beta_j,\theta} (\nu_{\rm rel}) \rangle
\\
&=\sum_{p,q}^{N_p} D_{p,i}^\theta H_{pq}^\theta D_{q,j}^\theta ,
\end{split}
\end{equation}
\begin{equation}
\begin{split}
  H_{pq}^\theta
&=\langle  \tilde  \phi_p | H^\theta| \phi_q \rangle
 =\langle  \tilde  \phi_p^{-\theta} | H | \phi_q^{-\theta} \rangle,
 \\
 D_{p,i}^\theta &= \langle  \tilde  \phi^{\beta_i,\theta}(\nu_{\rm rel}) | \phi_p \rangle = \langle \phi^{\beta_i} (\nu_{\rm rel}) | \phi_p^{-\theta} \rangle,
 \\
 N_{ij}^\theta&= \langle  \tilde  \phi^{\beta_i,\theta}(\nu_{\rm rel}) | \phi^{\beta_j,\theta}(\nu_{\rm rel}) \rangle
 =\sum_{p}^{N_p} D_{p,i}^\theta D_{p,j}^\theta. 
 \label{eq:csme}
\end{split}
\end{equation}
Here we insert the the completeness relation consisting of the states with a finite number $N_p$ ; $1=\sum_{p=1}^{N_p} | \phi_p \rangle \, \langle  \tilde  \phi_p|$.
In this study, we construct the completeness relation in terms of the nodeless HO basis function with $\theta=0$,
which are the same as those used in the projection operator method. We use the same set of $\{b_i\}$.
\begin{equation}
\begin{split}
  \phi_p (\vc{r}) &= \sum_{i=1}^{N_{\rm base}} C_{i,p}\, \phi_{0 \ell m }(\vc{r},b_i),
  \\
  \langle \tilde \phi_p | \phi_q \rangle&= \delta_{pq}.
  \\
\end{split}
\end{equation}
We diagonalize the norm matrix of $\phi_{0 \ell m }(\vc{r},b_i)$ and construct the orthonormalized basis states $\{\phi_p\}$ in the linear combination of $\phi_{0 \ell m }(\vc{r},b_i)$ with the coefficients $C_{i,p}$,
which nicely describe the completeness relation in the present calculation.
The states $\{\phi_p\}$ involve the Pauli-forbidden states, which are removed by diagonalizing the norm matrix with the elements of $N_{ij}^\theta$ in Eq.~(\ref{eq:csme}),
because of the overlap with the coherent basis state in $\{D_{p,i}^\theta\}$.
In the eigenvalue problem in Eq. (\ref{eq:eigen}),
when one diagonalizes the norm matrix, the eigenstates of the Pauli-forbidden states show the zero-energy eigenvalue,
which are removed from the basis states before diagonalizing the Hamiltonian matrix.
It is noted that this procedure is used for the calculation of the unbound states with the complex scaling only,
and is not necessary for the bound-state calculation with $\theta=0$.

%%%%%%%%%%%%%
\subsection{Level density}\label{sec:level}

In the complex scaling, the solutions $\{\Phi_n^\theta,\tilde \Phi_n^\theta\}$ construct the completeness relation \cite{berggren68,myo98} given as
\begin{equation}
  1  = \sum_{n} \kets{\Phi_n^\theta}\bras{\tilde \Phi_n^\theta},
          \label{eq:ECR}
\end{equation}
where $n$ is the state index. Using the energy eigenvalues $\{E_n^\theta\}$, the complex-scaled Green's function ${\cal G}^\theta(E)$ is expressed as
\begin{equation}
	{\cal G}^\theta(E)
=	\frac{ 1 }{ E- H^\theta }
=	\sum_n \frac{|\Phi^\theta_n\rangle \langle \tilde{\Phi}^\theta_n|}{E-E_n^\theta}.
	\label{eq:green1}
\end{equation}
We calculate the level density $\rho(E)=\sum_{n}\delta(E-E_n)$ with complex scaling \cite{suzuki05,odsuren14,odsuren21}.
The complex-scaled level density $\rho^\theta(E)$ is given with ${\cal G}^\theta(E)$ as
\begin{equation}
        \rho^\theta(E)
=     -\frac1{\pi} {\rm Im}\left\{{\rm Tr}\, {\cal G}^\theta(E) \right\} 
=     -\frac1{\pi} \sum_n {\rm Im}\left( \frac{1}{E-E_n^\theta} \right). 
        \label{eq:LD}
\end{equation}

We also consider the asymptotic Hamiltonian $H_0^\theta$ with the energy eigenvalues $\{E_{0,n}^\theta\}$,
and define the asymptotic level density $\rho_0^\theta(E)$ as
\begin{equation}
        \rho_0^\theta(E)
=       -\frac1{\pi}\ \sum_n {\rm Im}\Biggl( \frac{1}{E-E_{0,n}^\theta} \Biggr) . 
        \label{eq:LD0}
\end{equation}
One defines the continuum level density $\Delta(E)=\rho^\theta(E) - \rho_0^\theta(E)$,
which is related to the scattering matrix $S(E)$ \cite{levine69}:
\begin{equation}
  \Delta(E) =\frac1{2\pi} {\rm Im} \frac{{\rm d}}{{\rm d} E} {\rm ln}\bigl\{{\rm det}\, S(E)\bigr\} .
\end{equation}
In the single channel, $\Delta(E)$ becomes the derivative of the phase shift $\delta(E)$ and the phase shift is obtained as
\begin{equation}
\begin{split}
  \delta(E) &= \pi \int_{-\infty}^E \Delta (E') {\rm d}E'.
  \label{eq:phase}
\end{split}
\end{equation}

We define the asymptotic Hamiltonian $H_0$ for the $2\alpha$ system as \cite{kamimura77}
\begin{equation}
    H_0 =  T_{\rm rel} +  \frac{4e^2}{r}.
    \label{eq:Ham_asym}
\end{equation}
We omit the nuclear interaction, and replace the Coulomb interaction with the point type.
In the asymptotic wave function of $2\alpha$, one omits the antisymmetrization between the nucleons in the different $\alpha$ clusters \cite{kamimura77,myo23}.
This means no Pauli-forbidden state in the relative motion between $2\alpha$ and then we set $N=\ell$ with $n=0$ in the coherent basis method.
We also omit the projection operator in $H_0$ in the projection operator method.

%%%%%%%%%%%%%%%%%%%%%%%%%%%%%%%%%%%%%%%%%%%%
\section{Results}\label{sec:result}

\subsection{$\alpha$-$\alpha$ system}
\begin{figure}[t]
\centering
\includegraphics[width=8.0cm,clip]{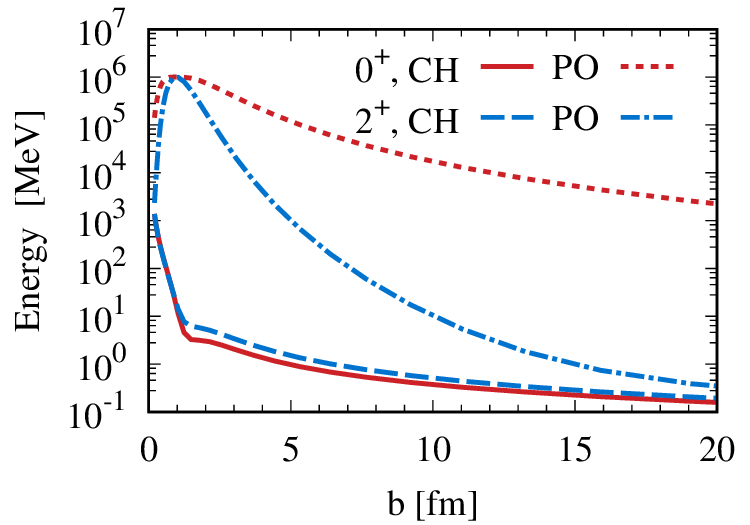}
\includegraphics[width=8.0cm,clip]{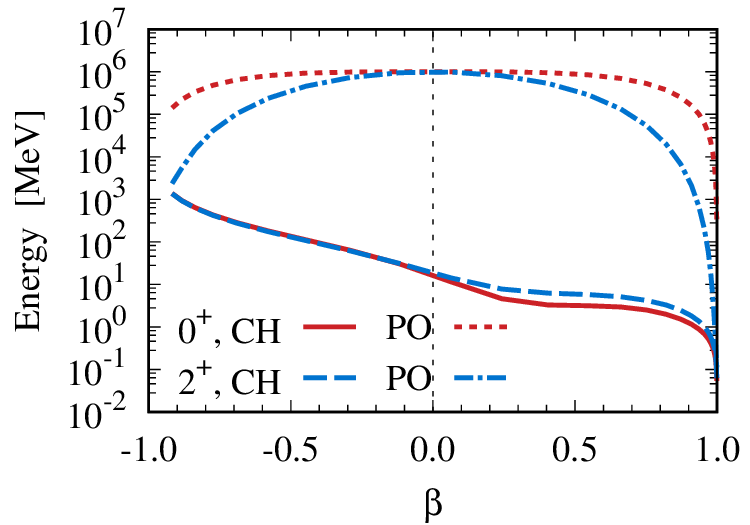}
\caption{Diagonal energies of the relative motion in $^8$Be for $0^+$ (red) and $2^+$ (blue) states
  as functions of the HO length parameter $b$ (top) and dilation parameter $\beta$ (bottom) in the coherent basis method (CH)
  and the projection operator method (PO) without complex scaling.}
\label{fig:ene_dia}
\end{figure}

In this study, we treat the $2\alpha$ system of $^8$Be and discuss the $\alpha$-$\alpha$ resonances.
First, we compare the diagonal energies of the basis states in the coherent basis method (CH) and the projection operator method (PO)
as functions of the HO length parameter $b$ in the Gaussians using Eq.~(\ref{eq:beta-b}).
In two methods, the treatments of Pauli-forbidden states are different, and affect the diagonal energies. 
We show the results of the $0^+$ and $2^+$ states in Fig.~\ref{fig:ene_dia} (top) in a logarithmic scale.
We also show the results as functions of the dilation parameter $\beta$ used in the coherent basis states in Fig.~\ref{fig:ene_dia} (bottom).
These figures are useful to understand the treatment of the Pauli principle in the coherent basis method, which leads to the low-energy states
in the large value of $b$, namely a large $\alpha$-$\alpha$ distance, and also in the values of $\beta$ close to unity.

In the projection operator method, the basis states in Eq.~(\ref{eq:WF_PO}) can involve the Pauli-forbidden states,
and then the pseudo potential with the strength of $\lambda=10^6$ MeV makes the states have high energies.
The HO length of Pauli-forbidden states is $b_{\rm rel}=$0.9667 fm and the maximum energies appear at this length for two spin states.
For the $0^+$ state, there are two Pauli-forbidden states with $n=0$ and $1$ and then
the repulsive effect is distributed in a wider range of $b$ than the results of the $2^+$ state, which includes one Pauli-forbidden state with $n=0$.
In the projection operator method, the superposition of the basis states makes the Pauli-allowed states with low energies.
The comparison of the two methods explains the reasonable treatment of the Pauli-allowed states in the coherent basis method. 

%%%%%%%%%%%%%%%%%%%%%%%%%%%%%%%%%%%%
%%%%%%%%%%%%%%%%%%%%%%%%%%%%%%%%%%%%

Next, we solve the eigenvalue problem of the Hamiltonian matrix.
For $0^+$ state, there are two Pauli-forbidden states and in the projection operator method, two states are obtained to have the high energies close to $\lambda$.
On the other hand in the coherent basis method, the basis states do not involve the Pauli-forbidden states,
and all eigenstates are obtained as the Pauli-allowed states.

Before the calculation of resonances, we discuss the reliability of the present coherent basis method for the bound state.
For this purpose, we artificially strengthen the $\alpha$-$\alpha$ nuclear potential $V^{\rm N}_{\alpha\alpha}(r)$ to make the $0^+$ and $2^+$ states of $^8$Be bound. 
We introduce the enhancement factor $\delta$ in $V^{\rm N}_{\alpha\alpha}(r)$ as
\begin{equation}
  V^{\rm N}_{\alpha\alpha}(r) \to V^{\rm N}_{\alpha\alpha}(r)(1+\delta). 
\end{equation}
We compare the resulting energies with those obtained in the projection operator method.

In Table \ref{tab:bound}, we show the energies of $^8$Be ($0^+$ and $2^+$) measured from the $\alpha$+$\alpha$ threshold energy by changing $\delta$.
It is found that the two methods give the same energies of the $0^+$ and $2^+$ states from weak to strong bindings with various values of $\delta$.
These results indicate the reliability of the present coherent basis method.

%%%%%%%%%%%%%%%%%%%%%%%%%%%%%% 
\begin{table}[b]
  \caption{Energies of $^8$Be ($0^+$ and $2^+$) measured from the $\alpha$+$\alpha$ threshold energy in MeV,
    calculated in two methods; coherent basis method (CH) and projection operator method (PO).
    The parameter $\delta$ is the enhancement factor of the $\alpha$-$\alpha$ nuclear potential.}
\label{tab:bound}
\centering
%\begin{ruledtabular}
%\small
\renewcommand{\arraystretch}{1.5}
\begin{tabular}{cllllll}
\noalign{\hrule height 0.5pt}
            &~~$0^+$~ &~~$0^+$~                        & & $2^+$&  $2^+$    & \\
~$\delta$~~ &~~CH~~ &~~PO~& & ~~CH~~ &~~PO~~\\
\noalign{\hrule height 0.5pt}                                                                      
$0.05$      &~~$-0.072$     &~~$-0.072$       & &  --       & --\\
$0.10$      &~~$-0.593$      &~~$-0.593$       & &  --       & --\\
$0.15$      &~~$-1.256$      &~~$-1.256$       & &  --       & -- \\
$0.20$      &~~$-2.065$      &~~$-2.065$       & &  --       & -- \\
$0.25$      &~~$-3.026$      &~~$-3.026$       & &  --       & -- \\
$0.30$      &~~$-4.147$      &~~$-4.147$       & & $-0.954$  & $-0.954$ \\
$0.35$      &~~$-5.430$      &~~$-5.430$       & & $-2.175$  & $-2.175$ \\
$0.40$      &~~$-6.880$      &~~$-6.880$       & & $-3.553$  & $-3.553$ \\
\noalign{\hrule height 0.5pt}
\end{tabular}
%\end{ruledtabular}
\end{table}
%%%%%%%%%%%%%%%%%%%%%%%%%%%%%%

\begin{figure}[t]
\centering
\includegraphics[width=8.0cm,clip]{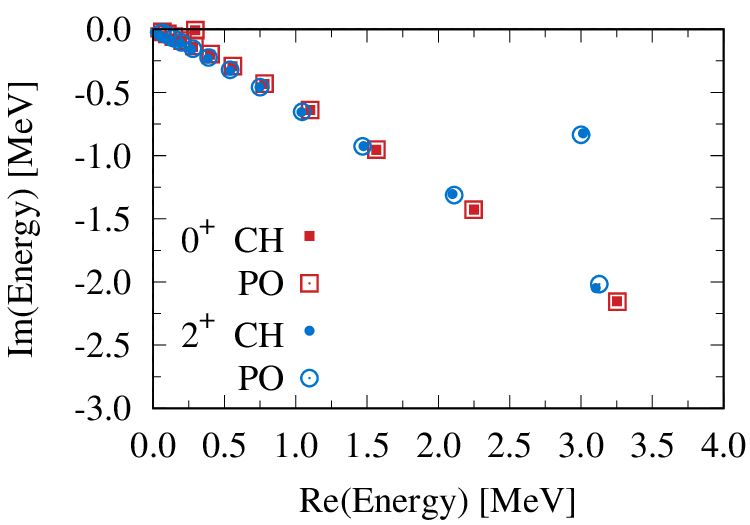}
\includegraphics[width=8.0cm,clip]{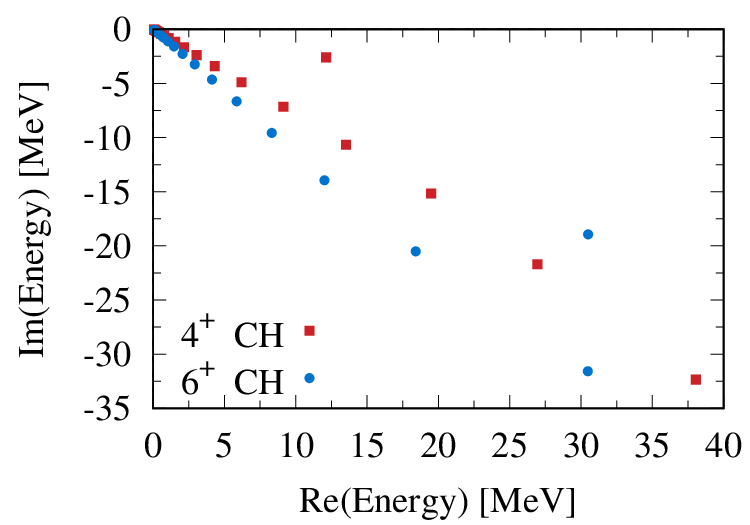}
\caption{Energy eigenvalues of $^8$Be (top: $0^+$ and $2^+$, bottom: $4^+$ and $6^+$) for the coherent basis method (CH, solid symbols) and the projection operator method (PO, open symbols) on the complex energy plane, measured from the $\alpha$+$\alpha$ threshold energy.
  Scaling angle $\theta$ is taken as 16$^\circ$ ($0^+$), 18$^\circ$ ($2^+$), 20$^\circ$ ($4^+$), and 25$^\circ$ ($6^+$).
The eigenvalues deviated from the line of discretized continuum states are resonances. 
}
\label{fig:ene_8Be}
\end{figure}

Next, we keep $\delta=0$ in the $\alpha$-$\alpha$ nuclear interaction and describe the unbound states of $^8$Be in the complex scaling.
We solve the complex-scaled eigenvalue problem in Eq.~(\ref{eq:eigen}) for 2$\alpha$ of $^8$Be ($0^+$, $2^+$, $4^+$, and $6^+$).
In Fig. \ref{fig:ene_8Be}, we show the energy eigenvalues $\{E_n^\theta\}$ of four spin states on the complex energy plane.
The scaling angle $\theta$ is optimized in each spin state from the stationary condition of
the energy eigenvalues of resonances on the complex energy plane with respect to $\theta$.
This condition gives $\theta=16^\circ, 18^\circ, 20^\circ$, and $25^\circ$ for $0^+, 2^+, 4^+$, and $6^+$, respectively. 
We show two kinds of solutions obtained in the coherent basis method (CH) and projection operator method (PO) in the $0^+$ and $2^+$ states.
For $4^+$ and $6^+$, the results obtained in the coherent basis method are shown.
The continuum states are discretized along a straight line and we obtain one resonance in each state deviating from the line of the continuum states.
In Fig. \ref{fig:ene_8Be}, the discretized continuum states also agree with each other by using the same range parameters in the relative wave function of $2\alpha$. 

In Table \ref{tab:energy}, we list the resonance energies and decay widths of four resonances of $^8$Be obtained in the coherent basis method
in comparison with the projection operator method. We also include the experimental data.
It is found that resonance energies and decay widths of two states of $^8$Be agree with each other in the two methods.
These results mean the reliability of the coherent basis method to describe resonances with complex scaling.

%%%%%%%%%%%%%%%%%%%%%%%%%%%%%% 
\begin{table}[b]
  \caption{Resonance parameters of $^8$Be measured from the $\alpha$+$\alpha$ threshold energy in MeV,
    in the coherent basis method (CH) and the projection operator method (PO).
    The experimental values (Exp.) are in the square brackets \cite{tilley04,nndc}.}
\label{tab:energy} 
\centering
%\begin{ruledtabular}
%\small
\renewcommand{\arraystretch}{1.5}
\begin{tabular}{ccclll}
\noalign{\hrule height 0.5pt}
~$J^\pm$~~&            & &~~energy~~        & &~~decay width~ \\
\noalign{\hrule height 0.5pt}                                                                       
          & CH   & &~ $~~~0.294$      & &~~$0.014$ \\  % th=16 Src03/Data02
$0^+$     & PO   & &~ $~~~0.296$      & &~~$0.015$ \\  % th=16
          & Exp. & &~ [0.0918]       & &~~[$5.57(25)\times 10^{-6}$]  \\
\hline
          & CH   & &~ $~~~3.01$       & &~~$~~1.65$       \\  % th=18 Src03/Data02
$2^+$     & PO   & &~ $~~~3.00$       & &~~$~~1.67$       \\  % th=18
          & Exp. & &~ [3.12(1)]      & &~~[1.513(15)]    \\
\hline
%          & CH   & &~ $~12.14$        & &~~$~~5.18$        \\ % th=20 Src03/Data02
\lw{$4^+$}& CH   & &~ $~12.13$        & &~~$~~5.19$        \\ % th=20 Src01/Data04
          & Exp. & &~ [11.44(15)]     & &~~[$\approx$ 3.5] \\
\hline
%          & CH   & &~ $~30.50$        & &~~$~38.09$       \\  % th=25 Src03/Data02
\lw{$6^+$}& CH   & &~ $~30.49$        & &~~$~37.88$       \\  % th=25 Src01/Data04
          & Exp. & &~ [$\approx$ 28]  & &~~[$\approx$ 20] \\
\noalign{\hrule height 0.5pt}
\end{tabular}
%\end{ruledtabular}
\end{table}
%%%%%%%%%%%%%%%%%%%%%%%%%%%%%%

%%%%%%%%%%%%%%%%%%%%%%%%%%%%%%%%%%%%
%%%%%%%%%%%%%%%%%%%%%%%%%%%%%%%%%%%%

\subsection{Phase shifts}
We calculate the eigenstates of the asymptotic Hamiltonian $H_0^\theta$ of $2\alpha$ using Eq.~(\ref{eq:Ham_asym})
to obtain the continuum level densities and phase shifts in the coherent basis method.
We employ the same set of the dilation parameters $\{\beta_i\}$ as used in the calculation with the full Hamiltonian $H^\theta$ and set the same scaling angle $\theta$ for each state.

\begin{figure}[t]
\centering
\includegraphics[width=8.0cm,clip]{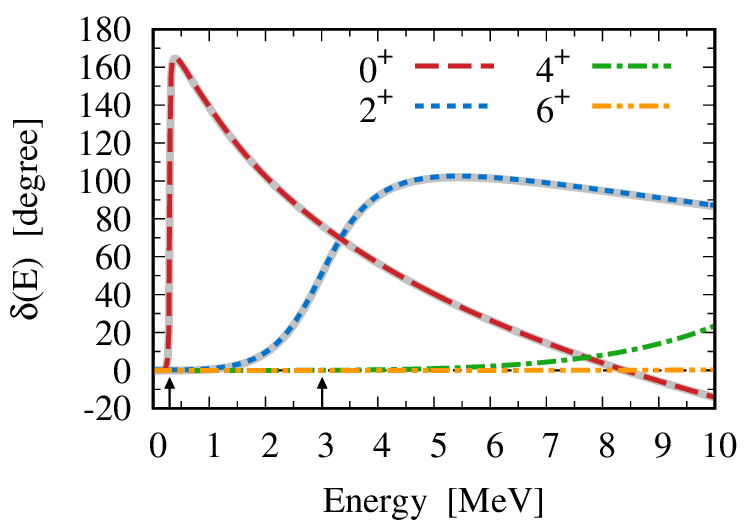}
\includegraphics[width=8.0cm,clip]{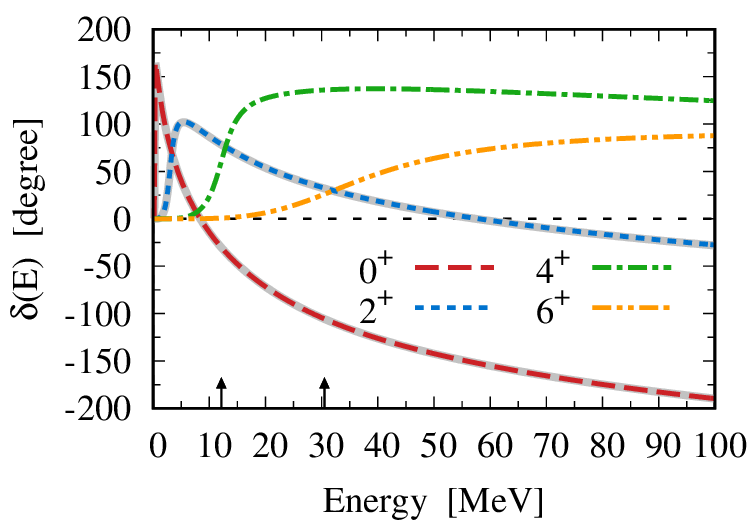}
\caption{Phase shifts of the $\alpha$--$\alpha$ scattering ($0^+$, $2^+$, $4^+$, and $6^+$) in the center-of-mass frame.
  The lines using dashed or dotted ones are the results in the coherent basis method and the gray solid lines for $0^+$, and $2^+$ are the ones in the projection operator method.
  The upper arrows from the bottom indicate the resonance energies of $0^+$, $2^+$, $4^+$, and $6^+$ in Table \ref{tab:energy}.}
\label{fig:PH_8Be}
\end{figure}

Using the energy eigenvalues $\{E_n^\theta\}$ and $\{E_{0,n}^\theta\}$ of 2$\alpha$, we calculate the continuum level density, $\Delta(E)$,
and evaluate the phase shift of the $\alpha$--$\alpha$ scattering by integrating $\Delta(E)$ in Eq. (\ref{eq:phase}).
In Fig.~\ref{fig:PH_8Be}, we show the phase shifts of the four states obtained, where we put the arrows at the resonance energies of four states shown in Table \ref{tab:energy}.

The resulting phase shifts with dashed or dotted lines are obtained in the coherent basis method,
and they agree with the gray lines obtained in the projection operator method for the $0^+$ and $2^+$ states.
In each state, the energy at the  maximum derivative of the phase shift is close to the resonance energy shown in the arrow.
From these results, one can apply the present coherent basis method to the scattering problem between various nuclear clusters with complex scaling.
One does not need the projection operator to eliminate the Pauli-forbidden states between clusters, which are automatically removed in the coherent basis method. 

%%%%%%%%%%%%%%%%%%%%%%%%%%%%%%%%%%%%
%\begin{figure*}[t]  % widefigure  
%\end{figure*}
%%%%%%%%%%%%%%%%%%%%%%%%%%%%%%%%%%%%

\section{Discussion}\label{sec:discuss}
We discuss the application of the present coherent basis method to the multicluster system beyond the two-cluster case.
We shall consider the $3\alpha$ system for $^{12}$C with two Jacobi coordinates of the $\alpha$-$\alpha$ and $2\alpha$-$\alpha$ systems.
We adopt the SU(3) representation for $^{12}$C with the coherent HO basis states \cite{yoshida11,kato88}, which is defined as
\begin{equation}
\begin{split}
  \Phi_{Q,(\lambda,\mu),JKM}^{\beta}
    &= \exp\left(\frac12 \beta \hat{D}^\dagger \right)\, \Phi_{Q,(\lambda,\mu),JKM},
    \\
    \hat{D}^\dagger
    &=\sum_{i=1}^2  \hat{D}_i^\dagger,
    \label{eq:three_base}
\end{split}
\end{equation}
where $i=1(2)$ is for the $\alpha$-$\alpha$ ($2\alpha$-$\alpha$) system with a quanta $N_i=2n_i+\ell_i$.
The total quanta of the basis state is given as $Q=N_1+N_2$ with the quanta of each Jacobi coordinate
under the irreducible SU(3) representation of $(\lambda,\mu)$ in the total spin $J$ with the $K$-quantum number.
The total raising operator $\hat{D}^\dagger$ is a summation of those for each Jacobi coordinate with the single dilation parameter $\beta$ in the exponent.
Using Eq.~(\ref{eq:three_base}), the basis state for the $3\alpha$ system is expressed
as the product of the relative wave functions with the coherent basis states $\Phi_{N_i\,\ell_i\,\gamma_i}^\beta$ as
%\begin{widetext}
\begin{equation}
\begin{split}
  \Phi^\beta_{Q,(\lambda,\mu)JKM}
  &= \sum_{N_1,N_2} C_{N_1 N_2 (\lambda,\mu)}
  \\
  &\times \sum_{\ell_1,\ell_2}
  \langle (N_1,0)\ell_1,(N_2,0)\ell_2||(\lambda,\mu)JK\rangle\,
  \\
  &\times
  \left[
  e^{\frac12 \beta \hat{D}^\dagger_1}\, \Phi_{N_1\,\ell_1\,\gamma_1},
  e^{\frac12 \beta \hat{D}^\dagger_2}\, \Phi_{N_2\,\ell_2\,\gamma_2}
  \right]_{JM}
  \\
  &= \sum_{N_1,N_2} C_{N_1 N_2 (\lambda,\mu)}
  \\
  &\times \sum_{\ell_1,\ell_2}
  \langle (N_1,0)\ell_1,(N_2,0)\ell_2||(\lambda,\mu)JK\rangle\,
  \\
  &\times
  \left[
  \Phi_{N_1\,\ell_1\,\gamma_1}^\beta,
  \Phi_{N_2\,\ell_2\,\gamma_2}^\beta
  \right]_{JM},
\end{split}
\end{equation}
%\end{widetext}
where $\gamma_1=2\nu$ for $\alpha$-$\alpha$ and $\gamma_2=8\nu/3$ for $2\alpha$-$\alpha$ are the HO range parameters in each relative motion,
and $\langle \cdots||\cdots \rangle$ is a SU(3) Clebsch Gordan coefficient.
The specific coefficient $C_{N_1 N_2 (\lambda,\mu)}$ is determined from the quanta in each relative motion and the $(\lambda,\mu)$ representation.
The total variational wave function is a superposition of the above basis states with various values of the quanta $Q$, $N_1$, $N_2$ with $(\lambda,\mu)$ and the dilation parameter $\beta$.
It is noted that the common $\beta$ is used in the two relative motions in the single basis state.
This condition comes to keep the symmetry of the identical $\alpha$ clusters,
which fixes the ratio of the range parameters of the coherent basis states for the two Jacobi coordinates to $\gamma_1/\gamma_2$.

We show the case of $Q=8$, $(\lambda,\mu)=(0,4)$, $J=0$ and $K=0$ for $^{12}$C, which uniquely gives $N_1=4$ and $N_2=4$, as
\begin{equation}
\begin{split}
  \Phi^\beta_{8,(0,4)}
  &=  \sum_{\ell_1=\ell_2=0,2,4}
  \langle (4,0)\ell_1,(4,0)\ell_2||(0,4)00 \rangle
  \\
  &\times
  \left[ \Phi_{4\,\ell_1\,\gamma_1}^\beta,
         \Phi_{4\,\ell_2\,\gamma_2}^\beta \right]_{00},
\end{split}
\end{equation}
\begin{equation}
\begin{split}
  \langle (4,0)0,(4,0)0||(0,4)00 \rangle &=\frac{8}{15},
  \\
  \langle (4,0)2,(4,0)2||(0,4)00 \rangle &=\frac{4}{3\sqrt{5}},
  \\
  \langle (4,0)4,(4,0)4||(0,4)00 \rangle &=\frac{3}{5}.
\end{split}
\end{equation}

We also define the basis states for the linear-chain states of $^{12}$C,
in which the lowest total-quanta is $Q=12$ with $(\lambda,\mu)=(12,0)$.
In this configuration, the sets of the quanta $(N_1,N_2)$ are given as (4, 8), (6, 6), (8, 4), and (10, 2).
In a similar way, extending the $3\alpha$ case,
the heavier multi-$\alpha$ cluster states can be constructed systematically in the SU(3) representation with the coherent basis states.
We plan to investigate the 3$\alpha$ structure in $^{12}$C in the present framework in the future.

%%%%%%%%%%%%%%%%%%%%%%%%%%%%%%%%%%%%%
\section{Summary}\label{sec:summary}
We presented a new scheme to construct the Pauli-allowed states in nuclei with the harmonic oscillator (HO) basis states.
We introduced a generalized coherent state of the HO basis state 
in terms of the raising operator $\hat{\vc{a}}^\dagger \cdot \hat{\vc{a}}^\dagger$ in the exponential form.
This basis state results in the HO basis state with the same quanta, but with the changeable range parameters, namely,
the radial dilation character. This property is important and controlled by one parameter, which we call the dilation parameter.
This coherent basis state is automatically orthogonal to the lower quanta state and 
represents the short-range and long-range properties of the particle motion from the dilation property of the basis state.
In this study, we utilized this property to treat the Pauli-allowed states appearing in relative motion of nuclear cluster systems.
We also extend this framework to treat the resonances and the cluster-cluster scattering in the complex scaling.

We show the application to the $2\alpha$ system of $^8$Be in the orthogonality condition model.
We compare the results in the coherent basis method with the conventional projection operator method,
in which the projection operator is imposed in the Hamiltonian to obtain the Pauli-allowed states.
It is confirmed that the present coherent basis method gives reasonable solutions of resonance energies, decay widths, and the phase shifts
of the $\alpha$-$\alpha$ scattering, which agree with those obtained in the projection operator method.
These results indicate the reliability of the coherent basis method.

We further discuss the extension of the present method to the multicluster systems and
explain the basic framework of the $3\alpha$ system of $^{12}$C.
We adopt the SU(3) representation of the HO basis states in the relative motions with the Jacobi coordinates 
and introduce the coherent basis states in each relative motion with a common dilation parameter. 
It would be interesting to apply this framework to investigate the multi-$\alpha$ cluster states of nuclei.

\section*{Acknowledgments}
This work was supported by JSPS KAKENHI Grants No. JP20K03962 and No. JP22K03643.

\appendix

\section{Generalized coherent state}\label{sec:appendix_A}

We formulate the generalized coherent state \cite{perelomov86} of the harmonic oscillator (HO) basis state
using the raising operator $\hat{D}^\dagger=\hat{\vc{a}}^\dagger \cdot \hat{\vc{a}}^\dagger$  \cite{rowe85}.
The HO basis state $\phi_{n\ell m}(\vc{r},\nu)$ with a range $\nu=1/b^2$ is usually defined
using the associated Laguerre polynomials $L_n^{(\ell+1/2)}(\nu r^2)$ as follows
\begin{equation}
  \begin{split}
    \phi_{n\ell m}(\vc{r},\nu) &= N_{n\ell}(\nu)\, e^{-\frac12 \nu r^2} L_n^{(\ell+1/2)}(\nu r^2)\, {\cal Y}_{\ell m}(\vc{r}),
    \\
    N_{n\,\ell}(\nu)
    & = \sqrt{\frac{\nu^{\ell+3/2}\, 2^{\ell+2}\, (2n)!!}{\sqrt{\pi}\, (2n+2\ell+1)!! }} ,
  \end{split}
\end{equation}
where $n$ represents the number of nodes in the radial wave function,
and $N=2n+\ell$ is a principal quantum number.
First, we start from the generating function for the associated Laguerre polynomials with $\alpha=\ell+1/2$ as 
\begin{equation}
\begin{split}
  \frac{e^{-\nu r^2\, t/(1-t)}}{(1-t)^{\alpha+1}} &= \sum_{m=0}^\infty L_m^{(\alpha)}(\nu r^2)\, t^m ,
\end{split}
\end{equation}
where $|t|<1$. We introduce the following $n$-th derivative of the generating function ${\cal S}_n$ with its expansion;
\begin{equation}
\begin{split}
    {\cal S}_n
  &= \frac{1}{n!} \frac{d^n}{d t^n} \left\{ \frac{e^{-\nu r^2\cdot t/(1-t)}}{(1-t)^{\alpha+1}}  \right\}
    \\
  &= \frac{1}{n!} \sum_{m=0}^\infty L_m^{(\alpha)}(\nu r^2)\, \frac{d^n t^m }{d t^n}
  \\
  &= \sum_{k=0}^\infty L_{n+k}^{(\alpha)}(\nu r^2)\, \frac{(n+k)!}{k!\, n!}\, t^{k},
 \label{eq:sn1}
\end{split}
\end{equation}
where $m=n+k$.
It is also proven that ${\cal S}_n$ is proportional to the associated Laguerre polynomials with the order of $n$
and the argument of $\nu r^2/(1-t)$ as;
\begin{equation}
\begin{split}
 {\cal S}_n
  &= \frac{e^{-\nu r^2\, t/(1-t)}}{(1-t)^{n+\alpha+1}} \, L_n^{(\alpha)}\left(\frac{\nu r^2 }{1-t} \right).
 \label{eq:sn2}
\end{split}
\end{equation}
This formula can be confirmed in the mathematical induction using 
the relation of ${\cal S}_{n+1}= (d {\cal S}_n /dt) \, /(n+1)$ and the properties of the associated Laguerre polynomials.
From two expressions of ${\cal S}_n$ in Eqs.~(\ref{eq:sn1}) and (\ref{eq:sn2}), we obtain the following relation
\begin{equation}
\begin{split}
  L_n^{(\alpha)}\left(\frac{\nu r^2}{1-t} \right)
  &=(1-t)^{n+\alpha+1}  \exp\left( \frac{t}{1-t} \nu r^2 \right)
  \\
  &\times
  \sum_{k=0}^\infty \frac{(n+k)!}{k!\, n! } t^k\, L_{n+k}^{(\alpha)}\left(\nu r^2 \right).
  \label{eq:laguerre}
\end{split}
\end{equation}
This formula means that the associated Laguerre polynomial with the argument of $\nu r^2/(1-t)$ and the order $n$ is expanded by $t^k$ in terms of those with the argument of $\nu r^2$ and the order of $n+k$.
This property is applicable to the HO basis states to connect the HO basis states with different range parameters.
Hereafter we define $\beta=-t$ for the dilation parameter in the coherent basis state
and use this relation in the HO basis state with the range $\nu/(1+\beta)$.

Next, we discuss the generalized coherent state with the dilation parameter $\beta$, which can be expanded in the HO basis states
using Eq.~(\ref{eq:HO_aa}), the quanta of which is larger than or equal to $N$,
because of the raising operator $\hat{D}^\dagger=\hat{\vc{a}}^\dagger \cdot \hat{\vc{a}}^\dagger$ as
\begin{equation}
\begin{split}
  \phi^\beta_{n \ell m}(\vc{r},\nu)
  &=\exp\left(\frac12 \beta \hat{D}^\dagger \right) \phi_{n \ell m}(\vc{r},\nu)
  \\
  &=\sum_{k=0}^\infty \frac{\beta^k}{2^k k!}\, 
  A_{n\ell} \left( \hat{\vc{a}}^\dagger \cdot \hat{\vc{a}}^\dagger \right)^{n+k} {\cal Y}_{\ell m}(\hat{\vc{a}}^\dagger)\,  \phi_{0}(\vc{r},\nu)
  \\
  &=\sum_{k=0}^\infty \frac{\beta^k}{2^k k!}\, 
  \frac{A_{n\,\ell}}{A_{n+k\,\ell}}\, \phi_{n+k\, \ell m}(\vc{r},\nu).
  \label{eq:coherent_expansion}
\end{split}
\end{equation}
We get the following relation for the ratio of the coefficients $A_{n\,\ell}/A_{n+k\,\ell}$ as
\begin{equation}
\begin{split}
  \frac{A_{n\,\ell}}{A_{n+k\,\ell}}
  &=(-1)^k \sqrt{
    \frac{ (2n+2k+2\ell+2)!\, (n+\ell+1)!\, (n+k)!}{(n+k+\ell+1)!\,(2n+2\ell+2)!\, n!} }.
\end{split}
\end{equation}
On the other hand, we define the following function $\varphi^\beta_{n \ell m}$ with a normalization constant $C$ and the HO basis state
with the range $\nu/(1+\beta)$,
\begin{equation}
\begin{split}
  \varphi^\beta_{n \ell m}(\vc{r},\nu)
   &=C\, \exp\left( \frac{\beta}{2(1+\beta)} \nu r^2 \right)\, \phi_{n \ell m}(\vc{r},\frac{\nu}{1+\beta}).
\end{split}
\end{equation}
Using Eq.~(\ref{eq:laguerre}) with $\beta=-t$,
\begin{equation}
\begin{split}
  \varphi^\beta_{n \ell m}(\vc{r},\nu)
  &=C\, \exp\left( -\frac{\nu (1-\beta)}{2(1+\beta)} r^2 \right) \, N_{n\ell}\left(\frac{\nu}{1+\beta}\right)
  \, {\cal Y}_{\ell m}(\vc{r})
  \\
  & \times
  (1+\beta)^{n+\ell+3/2}  \exp\left( -\frac{\beta}{1+\beta} \nu r^2 \right)
  \\
  & \times
  \sum_{k=0}^\infty \frac{(n+k)!}{k!\, n! } (-\beta)^k\, L_{n+k}^{(\ell+1/2)}\left(\nu r^2 \right)
  \\
  &=C\,   (1+\beta)^{n+\ell+3/2}
  \sum_{k=0}^\infty \frac{(n+k)!}{n! }\, \frac{(-\beta)^k}{k!} \, 
  \\
  & \times
  \frac{N_{n\ell}\left(\frac{\nu}{1+\beta}\right)}{N_{n+k\,\ell}(\nu)}\,  \phi_{n+k\, \ell m}(\vc{r},\nu).
\end{split}
\end{equation}
Here, 
\begin{equation}
\begin{split}
  \frac{N_{n\ell}\left(\frac{\nu}{1+\beta}\right)}{N_{n+k\,\ell}(\nu)}
  & = \sqrt{\frac{(2n)!!\, (2n+2k+2\ell+1)!!}{(1+\beta)^{\ell+3/2}\, (2n+2k)!!\, (2n+2\ell+1)!! }}
  \\
  & = \frac{1}{\sqrt{(1+\beta)^{\ell+3/2}}}\, \frac{n!\, (-1)^k }{2^k\,(n+k)!} \, \frac{A_{n\,\ell}}{A_{n+k\,\ell}}.
\end{split}
\end{equation}
Hence we can rewrite $\varphi^\beta_{n \ell m}$ using Eq. (\ref{eq:coherent_expansion}) as 
\begin{equation}
\begin{split}
 \varphi^\beta_{n \ell m}(\vc{r},\nu)
 &=C\, \sqrt{(1+\beta)^{2n+\ell+3/2}}\,\sum_{k=0}^\infty \frac{\beta^k}{2^k\, k!}\, \frac{A_{n\,\ell}}{A_{n+k\,\ell}}
 \\
  & \times
 \phi_{n+k\, \ell m}(\vc{r},\nu)
  \\
  &=C\, \sqrt{(1+\beta)^{\bar N}}\, \exp\left(\frac12 \beta \hat{D}^\dagger \right)\, \phi_{n \ell m}(\vc{r},\nu)
  \\
  &=C\, \sqrt{(1+\beta)^{\bar N}}\, \phi^\beta_{n \ell m}(\vc{r},\nu),
  \\
\end{split}
\end{equation}
where $\bar N=2n+\ell+3/2=N+3/2$.
Imposing the relation of $\varphi^\beta_{n \ell m}=\phi^\beta_{n \ell m}$, we can determine $C$ as
\begin{equation}
\begin{split}
  C&=\frac{1}{\sqrt{(1+\beta)^{\bar N}}}.
\end{split}
\end{equation}
Finally, we define the generalized coherent state of the HO basis state.
\begin{equation}
\begin{split}
 \phi^\beta_{n \ell m}(\vc{r},\nu)
  &=\exp\left(\frac12 \beta \hat{D}^\dagger \right)\, \phi_{n \ell m}(\vc{r},\nu)
   \\
  &=
   \frac{1}{\sqrt{(1+\beta)^{\bar N}}}\, \exp\left( \frac{\beta}{2(1+\beta)} \nu r^2 \right)
   \\
   & \times
   \phi_{n \ell m}(\vc{r},\frac{\nu}{1+\beta}).
\end{split}
\end{equation}
It is noted that the above coherent basis state is not normalized, and one can normalize it in the calculation of the norm matrix element.
\section{Kinetic energy}\label{sec:appendix_B}

We give the formula of the matrix element of the kinetic energy $T$ with a reduced mass $\mu$ in the generalized coherent basis states
with the independent values of $\beta$ and $n$ in the bra and ket states.
\begin{equation}
\begin{split}
  \langle \phi_{n'\ell}^{\beta'}(\nu) |T| \phi_{n\ell}^{\beta}(\nu) \rangle
  &=  \frac{\hbar^2\nu_{\beta}}{2\mu}\,  f_{n'n\,\ell}^{\beta'\beta}(\nu),
  \\
  f_{n'n\, \ell}^{\beta'\beta}(\nu)
  &=
  \left( 1 - \beta \right)^2\, A_{n}\, G_{n'n+1\,\ell}^{\beta'\beta}(\nu)
  \\
  & + \left (1-\beta \right)  \left (1+\beta \right)\, B_{n}\, G_{n'n\,\ell}^{\beta'\beta}(\nu)
  \\
  & + \left( 1 + \beta \right)^2\, C_{n}\, G_{n'n-1\,\ell}^{\beta'\beta}(\nu),
\end{split}
\end{equation}
where
\begin{equation}
\begin{split}
  G_{n'n\,\ell}^{\beta'\beta}(\nu) &=
  \frac{ \langle \phi_{n'\ell}(\nu_{\beta'}) | e^{\left( \gamma_{\beta'}  + \gamma_{\beta} \right) r^2} |\phi_{n \ell}(\nu_\beta)\rangle}{\sqrt{(1+\beta')^{\bar N'}(1+\beta)^{\bar N}}},
  \\
  A_{n}&=\sqrt{(n+1)(n+\ell+\frac32)},
  \\
  B_{n}&= 2n+\ell+\frac32 ,\quad
  C_{n} =\sqrt{n(n+\ell+\frac12)},
  \\
  \nu_\beta &=\frac{\nu}{1+\beta},\qquad
  \gamma_\beta =  \frac{\beta }{1+\beta }\, \frac{\nu}{2}.
  \\
\end{split}
\end{equation}
\hspace*{8cm}
\section*{References}
%\vfill\pagebreak
%%%%%%%%%%%%%%%%%%%%%%%%%%%%%%%%%%%%%%%%%%%%%%%%%%%%%%%%%%%%%
%%%%%%%%%%%%%%%%%%%%%%%%%%%%%%%%%%%%%%%%%%%%%%%%%%%%%%%%%%%%%
\def\JL#1#2#3#4{ {{\rm #1}} \textbf{#2}, #4 (#3)}  % Physical Review (year last)
\nc{\PR}[3]     {\JL{Phys. Rev.}{#1}{#2}{#3}}
\nc{\PRC}[3]    {\JL{Phys. Rev.~C}{#1}{#2}{#3}}
\nc{\PRA}[3]    {\JL{Phys. Rev.~A}{#1}{#2}{#3}}
\nc{\PRL}[3]    {\JL{Phys. Rev. Lett.}{#1}{#2}{#3}}
\nc{\NP}[3]     {\JL{Nucl. Phys.}{#1}{#2}{#3}}
\nc{\NPA}[3]    {\JL{Nucl. Phys. A}{#1}{#2}{#3}}
\nc{\PL}[3]     {\JL{Phys. Lett.}{#1}{#2}{#3}}
\nc{\PLB}[3]    {\JL{Phys. Lett.~B}{#1}{#2}{#3}}
\nc{\PTP}[3]    {\JL{Prog. Theor. Phys.}{#1}{#2}{#3}}
\nc{\PTPS}[3]   {\JL{Prog. Theor. Phys. Suppl.}{#1}{#2}{#3}}
\nc{\PTEP}[3]   {\JL{Prog. Theor. Exp. Phys.}{#1}{#2}{#3}}
\nc{\PRep}[3]   {\JL{Phys. Rep.}{#1}{#2}{#3}}
\nc{\PPNP}[3]   {\JL{Prog.\ Part.\ Nucl.\ Phys.}{#1}{#2}{#3}}
\nc{\JPG}[3]     {\JL{J. of Phys. G}{#1}{#2}{#3}}
\nc{\andvol}[3] {{\it ibid.}\JL{}{#1}{#2}{#3}}
%%%%%%%%%%%%%%%%%%%%%%%%%%%%%%%%%%%%%%%%%%%%%%%%%%%%%%%%%%%%%

\end{document}